\documentclass[usenatbib]{mn2e}

\usepackage{psfig,graphicx}

\def\chisq{\hbox{$\chi^2$}}
\def\chinu{\hbox{$\chi^2_\nu$}}
\def\msun{\hbox{${\rm M}_{\odot}$}}
\def\mspy{\hbox{${\rm M}_{\odot}$\,yr$^{-1}$}}
\def\rsun{\hbox{${\rm R}_{\odot}$}}

\def\rstar{\hbox{$R_{\star}$}}
\def\teff{\hbox{$T_{\rm eff}$}}
\def\logg{\hbox{$\log g$}}
\def\sn{\hbox{S/N}}

\def\kms{\hbox{km\,s$^{-1}$}}
\def\vsini{\hbox{$v\sin(i)$}}

\def\ptt{\hbox{$10^{-4} I_{\rm c}$}}

\def\degr{\hbox{$^\circ$}}

\def\prot{\hbox{$P_{\rm rot}$}}
\def\vinf{\hbox{$v_{\infty}$}}
\def\Mdot{\hbox{$\dot{M}$}}
\def\zetaori{\hbox{$\zeta$~Ori~A}}
\def\tori{\hbox{$\theta^1$~Ori~C}}
\def\vturb{\hbox{$v_{\rm turb}$}}
\def\vmax{\hbox{$v_{\rm max}$}}
\def\vmin{\hbox{$v_{\rm min}$}}
\def\vcl{\hbox{$v_{\rm cl}$}}
\def\logLX{\hbox{$\log L_{\rm X}/L_{\rm bol}$}}

\begin{document}

\title[The weak magnetic field of $\zeta$~Ori A] {The weak magnetic 
field of the O9.7 supergiant $\zeta$~Orionis A\thanks{Based on observations obtained at the
T\'elescope Bernard Lyot (TBL), operated by the Institut National des Science de 
l'Univers of the Centre National de la Recherche Scientifique of France.} }

\makeatletter

\def\newauthor{%
  \end{author@tabular}\par
  \begin{author@tabular}[t]{@{}l@{}}}
\makeatother
 
\author[Bouret et al]
{\vspace{1.5mm}
J.-C.~Bouret$^1$, J.-F.~Donati$^2$, F.~Martins$^3$, C.~Escolano$^1$, W.~Marcolino$^1$, \\ 
\vspace{1.7mm}
{\hspace{-1.5mm}\LARGE\rm
T.~Lanz$^4$, I.D.~Howarth$^5$ }\\
$^1$ LAM--UMR 6110, CNRS \& Univ.\ de Provence, 38 rue Fr\'ed\'eric Joliot-Curie, F--13388 Marseille cedex 13, France\\
$^2$ LATT--UMR 5572, CNRS \& Univ.\ de Toulouse, 14 Av.\ E.~Belin, F--31400 Toulouse, France\\
$^3$ GRAAL--UMR 5024, CNRS \& Univ.\ de Montpellier II, Place Bataillon, F--34095 Montpellier, France\\ 
$^4$ Department of Astronomy, University of Maryland, College Park, MD 20742, USA\\  
$^5$ Dept.\ of Physics and Astronomy, University College London,  Gower Street, London WC1E6BT, UK\\ 
}

\date{Accepted, 2008 June 12}
\maketitle
 
\begin{abstract}  
We report here the detection of a weak magnetic field of 50--100~G on the O9.7 
supergiant $\zeta$~Ori A, using spectropolarimetric observations obtained with NARVAL 
at the 2m T\'elescope Bernard Lyot atop Pic du Midi (France).  
\zetaori\ is the third O star known to host a magnetic field (along with \tori\ and HD~191612), and
the first detection on a 'normal' rapidly-rotating O star. The magnetic field of \zetaori\ is the weakest magnetic field ever detected on a massive star. The measured field is lower than the thermal equipartition limit (about 100~G).
By fitting NLTE model atmospheres to our spectra, we determined that \zetaori\ 
is a 40~\msun\ star with a radius of 25~\rsun and an age of about 5--6~Myr, 
showing no surface nitrogen enhancement and losing mass at a rate of about 2$\times10^{-6}$~\mspy. 

The magnetic topology of \zetaori\ is apparently more complex than a dipole and 
involves two main magnetic polarities located on both sides of the same hemisphere;  
our data also suggest that \zetaori\ rotates in about 7.0~d and is about 40\degr\ away 
from pole-on to an Earth-based observer.  Despite its weakness, the detected magnetic 
field significantly affects the wind structure;  the corresponding Alfv\'en 
radius is however very close to the surface, thus generating a different rotational 
modulation in wind lines than that reported on the two other known magnetic O stars.  

The rapid rotation of \zetaori\ with respect to \tori\ appears as a surprise, both 
stars having similar unsigned magnetic fluxes (once rescaled to the same radius);  
it may suggest that the sub-equipartition field detected on \zetaori\ is not a 
fossil remnant (as opposed to that of \tori\ and HD~191612), but the result of an 
exotic dynamo action produced through MHD instabilities.   
\end{abstract}


\section{Introduction}
Stellar magnetic fields have been detected across a large range of spectral types. 
In solar-type and essentially all cool, low-mass (i.e., mid F and later) stars, 
magnetic fields (and activity) are observed, often featuring a complex topology, and 
are thought to be due to dynamo processes occuring within the outer convective layers.  
In hotter, more massive stars with outer radiative zones, magnetic fields are also 
detected (with a significantly simpler topology though) but only in a small fraction 
of stars (e.g., the magnetic chemically peculiar stars among the A and 
late B stars).  The situation might be similar (though less well studied) among 
early B and O stars, with only two O stars (namely \tori\ and  HD~191612, 
\citealt{Donati02, Donati06a}) and less than a handfull of early B-type stars (e.g., 
$\tau$~Sco, $\beta$~Cep, $\zeta$~Cas, \citealt{Donati06b, Donati01, Neiner03a}) yet 
known as magnetic.  

Magnetic fields are nonetheless expected to play a significant role throughout the 
evolution of hot massive stars, by modifying the internal rotation, enhancing 
chemical transport and mixing, and producing enhanced surface abundances \citep{Maeder03, 
Maeder04, Maeder05}.  Magnetic fields can also dramatically influence 
the way winds are launched \citep[e.g.,][]{udDoula02} and the later phases of evolution 
\citep[e.g., the collapse, ][]{Heger05};  a large number of
observational phenomena (e.g., non-thermal radio emission, anomalous X-ray spectra, 
abundance anomalies, and H$\alpha$ modulation) can also be explained (qualitatively at 
least) by the existence of a weak magnetic field.
Yet, the origin of magnetism in massive stars is still an open question, with a lively debate
between two classes of models.
%Yet, the origin of magnetism in massive stars is still a matter of speculation, with two 
%classes of models confronting to each other. 
While some models assert that dynamo processes 
(either located in the convective core, e.g., \citealt{Charbonneau01}, or acting within 
the radiative zone, e.g., \citealt{Mullan05}) can produce the observed magnetic fields, 
some others claim that the field is fossil in nature \citep{Ferrario05, Ferrario06}, being 
advected and amplified through the initial protostellar collapse.  

The limited knowledge that we have about the existence and statistical properties of magnetic 
fields in massive O stars is mostly due to the fact that these fields are difficult to 
detect.  Absorption lines of O stars are both relatively few in number in the optical domain, 
and generally rather broad (because of rotation or to some other type of as yet unknown 
macroscopic mechanism, e.g., \citealt{Howarth97}), decreasing dramatically the size of 
the Zeeman signatures that their putative fields can induce.
The results obtained so far (on two stars only) suggest that magnetic O-type stars may be 
(i) slow rotators and (ii) may exhibit a peculiar spectrum with very regular temporal 
modulation.   
While this view may partly reflect an observational bias (magnetic detections being easier on 
slow rotators) or a selection effect (observations often concentrating on peculiar stars 
first), null results recently reported on intermediate and fast rotators argue that this 
effect may be real.  This question is nevertheless a key point for clarifying both the origin 
and evolutionary impact of magnetic fields in massive stars and therefore deserves being 
studied with great care.  

With the advent of the new generation spectropolarimeters, such as ESPaDOnS at the 
Canada-France-Hawaii Telescope (CFHT) in Hawaii and NARVAL on the T\'elescope Bernard 
Lyot (TBL) in southern France, studies of stellar magnetic fields have undergone a big 
surge of activity;  in particular, detecting magnetic fields of massive O stars (or 
providing upper limits of no more than a few tens of G) is now within reach.  
In this context, we recently initiated a search for magnetic fields in a limited number 
of 'normal' O stars, using NARVAL.  

One of our targets is \zetaori, a  O9.7~Ib supergiant 
\citep{ma04} and the brightest O star at optical wavelengths.  
Evidence for azimuthal wind structuration (with an modulation timescale of about 6~d, 
compatible with the rotation period) is reported from both UV and optical lines 
\citep[e.g.,][]{Kaper96, Kaper99} and possibly due to the presence of a weak magnetic field.  
\zetaori\ is also well-known for its prominent X-ray emission, $\logLX=-6.74$  \citep{berghofer97}.  The origin of this X-ray emission 
is however still controversial;  while \citet{Cohen06} suggest that it is due to the classical 
wind-shock mechanism (with X-rays originating from cooling shocks in the acceleration zone), 
\citet{Raassen08} invoke a collisional ionization equilibrium model and \citet{Pollock07} 
argue for collisionless shocks controlled by magnetic fields in the wind terminal velocity 
regime.  For all these reasons, \zetaori\ is an obvious candidate for our magnetic 
exploration program.  

In this paper, we report our spectropolarimetric observations of \zetaori\ and 
present the Zeeman detections we obtained (Sec.~\ref{sec:obs}).  From the collected 
spectra, we reexamine the fundamental parameters of \zetaori\ and discuss the observed 
rotational modulation to attempt pinning down the rotation period (Sec.~\ref{sec:spec}).  
We then carry out a complete modeling of the detected Zeeman signatures and describe 
the reconstructed magnetic topology (Sec.~\ref{sec:magt}).  We finally summarise our 
results, discuss their implications for our understanding of massive magnetic stars, 
and suggest new observations to confirm and expand our conclusions (Sec.~\ref{sec:disc}).

\section{Observations}
\label{sec:obs}

Spectropolarimetric observations of \zetaori\ were collected with 
NARVAL at TBL in 2007 October, as part of a 
10-night run aimed at investigating the magnetic fields of hot stars.  
%(Results obtained on the other stars observed in the same run will 
%be published separately.)  
\zetaori\ was observed during seven nights;  altogether, 
292 circular-polarization sequences, each consisting of four individual
subexposures taken in different polarimeter configurations, were
obtained.  
From each set of four subexposures we derive a mean Stokes $V$ spectrum
following the procedure of \citet{Donati97}, ensuring in particular that all
spurious signatures are removed at first order.
Null polarisation spectra (labelled $N$) are calculated by combining the four 
subexposures in such a way that polarization cancels out, allowing us 
to check that no spurious signals are present in the data
(\citealt[see]{Donati97} for more details on how $N$ is defined).
All frames were
processed using Libre~ESpRIT (\citealt{Donati97}; Donati et al., in
prep.), a fully automatic reduction package installed at TBL for
optimal extraction of NARVAL spectra.  The peak signal-to-noise
ratios per 2.6~\kms\ velocity bin range from 800 to 1500, depending
mostly on weather conditions (see Table~\ref{tab:log}).

\begin{table*}
\caption[]{Journal of observations.  Columns 1--5 list the 
date, the range of heliocentric Julian dates, the range of UT times, 
the number of sequences and the exposure time per individual sequence 
subexposure, and the range of peak signal to noise ratio 
(per 2.6~\kms\ velocity bin), for each night of observation.  
Column 6 lists the rms noise level (relative to
the unpolarized continuum level and per 7.2~\kms\ velocity bin) in
the circular polarization profile produced by Least-Squares
Deconvolution once averaged over the whole night (Section~\ref{sec:obs}). 
The rotation cycle (using the ephemeris given by Eq.~\ref{eq:eph}) 
is listed in column 7. }
\begin{tabular}{ccccccc}
\hline 
Date & HJD          & UT      & $t_{\rm exp}$ & \sn\  & $\sigma_{\rm LSD}$ & Phase \\
(2007)     & (2,454,000+) & (h:m:s) &   (s)         &       &   (\ptt) &  \\
\hline
Oct.\ 18 & 391.53928--391.69519 & 00:52:56--04:37:25 & $48\times4\times20$ &  780-- 990 & 0.41 & 0.648--0.670 \\
Oct.\ 19 & 392.69363--392.72343 & 04:35:04--05:17:59 & $ 8\times4\times40$ & 1010--1080 & 0.87 & 0.813--0.817 \\
Oct.\ 20 & 393.54410--393.72674 & 00:59:39--05:22:38 & $44\times4\times40$ & 1220--1470 & 0.28 & 0.935--0.961 \\
Oct.\ 21 & 394.46543--394.66510 & 23:06:15--03:53:46 & $48\times4\times40$ &  810--1460 & 0.28 & 1.066--1.095 \\
Oct.\ 22 & 395.49302--395.69959 & 23:45:52--04:43:19 & $48\times4\times40$ & 1090--1480 & 0.27 & 1.213--1.242 \\
Oct.\ 24 & 397.47086--397.67036 & 23:13:45--04:01:01 & $48\times4\times40$ & 1030--1480 & 0.27 & 1.495--1.524 \\
Oct.\ 25 & 398.50205--398.70310 & 23:58:34--04:48:03 & $48\times4\times40$ & 1200--1470 & 0.27 & 1.643--1.671 \\
\hline
\end{tabular}
\label{tab:log}
\end{table*}

Least-Squares Deconvolution (LSD; \citealt{Donati97}) was applied to
all observations.  The line list was constructed manually to include 
the few moderate to strong absorption lines that are not (or only 
weakly) affected by the wind.  The strong Balmer lines, all showing 
clear emission from the wind and/or circumstellar environment at the 
time of our observations, were also excluded from the list.  
The C~{\sc iv} lines at 580.13 and 581.20~nm are used as reference 
photospheric lines from which we obtain the average radial velocity 
of \zetaori\ (about 45~\kms); a few 
unblended absorption lines that are not blueshifted with respect to the 
reference frame by more than 15~\kms\ are also included in the list.  
We end up with a list of only six lines, whose characteristics are summarised in 
Table~\ref{tab:lines}.

\begin{table}
\caption[]{Lines used for Least-Squares Deconvolution.  The
line depths (column 3) were directly measured from our spectra while
the Land\'e factors (column 4) were derived assuming LS coupling. } 
\begin{tabular}{clcc}
\hline
Wavelength & Element & Depth & Land\'e   \\
(nm)       &         & ($I_{\rm c}$) & factor      \\
\hline
492.1931   & $\qquad$He {\sc i}  & 0.44 & 1.000 \\
501.5678   & $\qquad$He {\sc i}  & 0.37 & 1.000 \\
541.1516   & $\qquad$He {\sc ii} & 0.35 & 1.000 \\
559.2252   & $\qquad$O {\sc iii} & 0.40 & 1.000 \\ 
580.1313   & $\qquad$C {\sc iv}  & 0.20 & 1.167 \\
581.1970   & $\qquad$C {\sc iv}  & 0.15 & 1.333 \\
\hline
\end{tabular}
\label{tab:lines}
\end{table}

From those lines we produced a mean circular polarization profile (LSD
Stokes $V$ profile), a mean check ($N$ for null) profile and a mean 
unpolarized profile (LSD Stokes $I$ profile) for each spectrum.  
All LSD profiles were produced on a spectral grid with a velocity bin 
of 7.2~\kms.  
Averaging together all LSD profiles recorded on each night of the seven nights 
of observation 
(with weights proportional to the inverse variance of each profile) yields 
relative noise levels of 0.27 (in units of \ptt) except on the first two 
nights (where the noise reaches 0.41 and 0.87).  
On Oct.~24, the detection probability exceeds 99\%, with a 
reduced-\chisq\ value (compared to a null-field, $V=0$ profile) of 1.33; 
the corresponding Stokes $V$ (and null $N$) LSD profiles are shown in 
Fig.~\ref{fig:lsd}. Similar (though less clear) Zeeman signatures are 
also observed during the other nights.

\begin{figure}
\center{\includegraphics[scale=0.35,angle=-90]{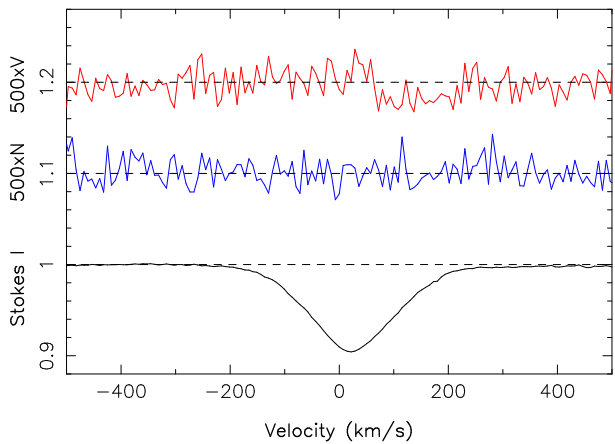}}
\caption[]{LSD Stokes $V$ (top), null $N$ (middle) and Stokes $I$ profiles of 
\zetaori\ on 2007 Oct~24.  The $V$ and $N$ profiles are expanded by a factor of
500 and shifted upwards by 1.2 and 1.1 for display purposes.  A clear 
Zeeman signature is detected in the red line wing while the null profile 
shows no signal.  }
\label{fig:lsd}
\end{figure}

\section{Parameters and rotation of \zetaori}
\label{sec:spec}

\subsection{The photosphere and wind of \zetaori}

We reexamine the spectral properties and fundamental parameters 
of \zetaori, using the new recorded spectra.  
%Unified models, with a consistent treatment of the photosphere and the wind, are mandatory to 
%analyze the spectra of late-type O supergiants. This is obviously true for P~Cygni profiles 
%and/or emission line profiles formed in the wind. This also applies to the bulk of spectral 
%lines formed in the photosphere, which could be severely contaminated by wind emission. In 
%addition, these photospheric lines are influenced by the velocity field in the transition 
%zone between the photosphere and the wind \citep{schaerer94}.

We have performed the spectrum analysis using model atmospheres calculated with the unified model 
code CMFGEN \citep{hillier98}. CMFGEN provides a consistent treatment of the photosphere and the wind, thus offering useful insights into the wind properties while providing a realistic treatment of photospheric metal-line blanketing.  
The code solves for the atmospheric structure, non-LTE 
populations and the radiation field, in the comoving frame of the fluid. The fundamental 
stellar parameters (\teff, \logg, \rstar\ and abundances) must be specified at this step, together 
with the mass-loss rate and velocity law.  After convergence of the model, a formal solution 
of the radiative transfer equation is computed  in the observer's frame, thus providing the 
synthetic spectrum for comparison to observations. 
For more details on this code, we refer to \citet{hillier98} and \citet{hillier03}.  
CMFGEN does not solve the full hydrodynamics, but rather assumes a density structure.  
We use a hydrostatic density structure computed with TLUSTY \citep{hubeny95, lanz03} in 
the deeper layers, while the wind regime is described with a standard $\beta$-velocity law.   
The photosphere and the wind are connected below the sonic point at a wind velocity of 
about 15~\kms. 

Radiatively driven winds are intrinsically subject to instabilities, resulting in the formation of 
discrete structures called ``clumps''. Both observational evidence and theoretical arguments 
foster the concept of highly-structured winds \citep{eversberg98,dessart03, dessart05}. To investigate spectral signatures of clumping in 
the wind of \zetaori\, and its consequences on the derived wind parameters, we have 
constructed clumped wind models with CMFGEN. A simple, parametric treatment of wind clumping is 
implemented in CMFGEN, which is expressed by a volume filling factor, $f$. It assumes a 
void interclump medium and the clumps to be small compared to the photons mean free path.  
Clumps start to form in the wind at velocities higher than $v_{\rm cl}$. We refer to \citet{hillier03}
for a detailed description of wind-clumping. 

%The filling factor is such that $\bar{\rho} = f\rho$, where $\bar{\rho}$ is the homogeneous 
%(unclumped) wind density. In other words, the density in the clumps is enhanced by a factor of f$^{-1}$ compared to the density  of a smooth model with the same mass-loss rate $\dot{M}$.
%Consequently, for density-squared diagnostics such as H$\alpha$, $\dot{M}$ in a smooth-wind model is overestimated by 1/$\sqrt{\rm f}$ (i.e. $\dot{M}_{c} = \sqrt{\rm f}\dot{M}_{s}$, where the $s$ and $c$
%subscripts refer to smooth and clumped models respectively).
%Besides, we assume that the filling factor decreases exponentially 
%with increasing radius (or, equivalently, with increasing velocity), following:  
%\begin{equation}
%f = f_\infty + (1-f_\infty) \exp(-v/v_{\rm cl}),
%\end{equation}
%where $v_{\rm cl}$ is the velocity at which clumping starts. 
We adopted a value of the clumping filling factor of $f=0.1$. Test models using $f=1$ revealed little change of the H$\alpha$ profile. As for $v_{\rm cl}$ , models with values from 30 \citep[see][]{bouret05} 
to 400~\kms\ showed a larger and larger shift of the central absorption component of H$\alpha$ towards shorter wavelengths. The best match was obtained for $v_{\rm cl} \sim 200$ \kms.
%{\bf We first ran several test models with homogeneous winds (i.e. $f=1$); these tests convincingly showed that homogeneous models are ruled out as they never provide a good fit to H$\alpha$
%(neither its core nore its wings). We thus turned to clumped models.
%We allowed $v_{\rm cl}$ to vary from 30~\kms\ (just above the sonic point, e.g. \citealt{bouret05}) to values of a few hundred \kms. We have adopted the standard value, $f=0.1$, because H$\alpha$ shows limited sensitivity to the filling factor. }

A depth-independent microturbulent velocity is included in the computation of the atmospheric structure 
(i.e., temperature structure \& population of individual levels). We chose a value of 5~\kms\ as the default 
value \citep[see][]{msh02}.
For the computation of the detailed spectrum resulting from a formal solution of the 
radiative transfer equation (i.e., with the populations kept fixed), a depth dependent microturbulent 
velocity was adopted. In that case, the microturbulent velocity follows the relation 
$\vturb(r) = \vmin + (\vmax - \vmin)\ v(r) / \vinf $ 
where  \vmin\ and \vmax\ are the minimum and maximum microturbulent velocities, 
and \vinf\ the terminal wind velocity.  
In this formulation, \vmin\ is technically equivalent to $\xi_{t}$, the microturbulence in
the photosphere. We considered several values of \vmin, searching for consistent fits for the photospheric lines. This is obtained for $\vmin=10$~\kms\ in the photosphere. $\vmax=0.1\ \vinf$ was adopted at the top of the atmosphere.

\begin{figure*}
\center{\hbox{\includegraphics[scale=0.45]{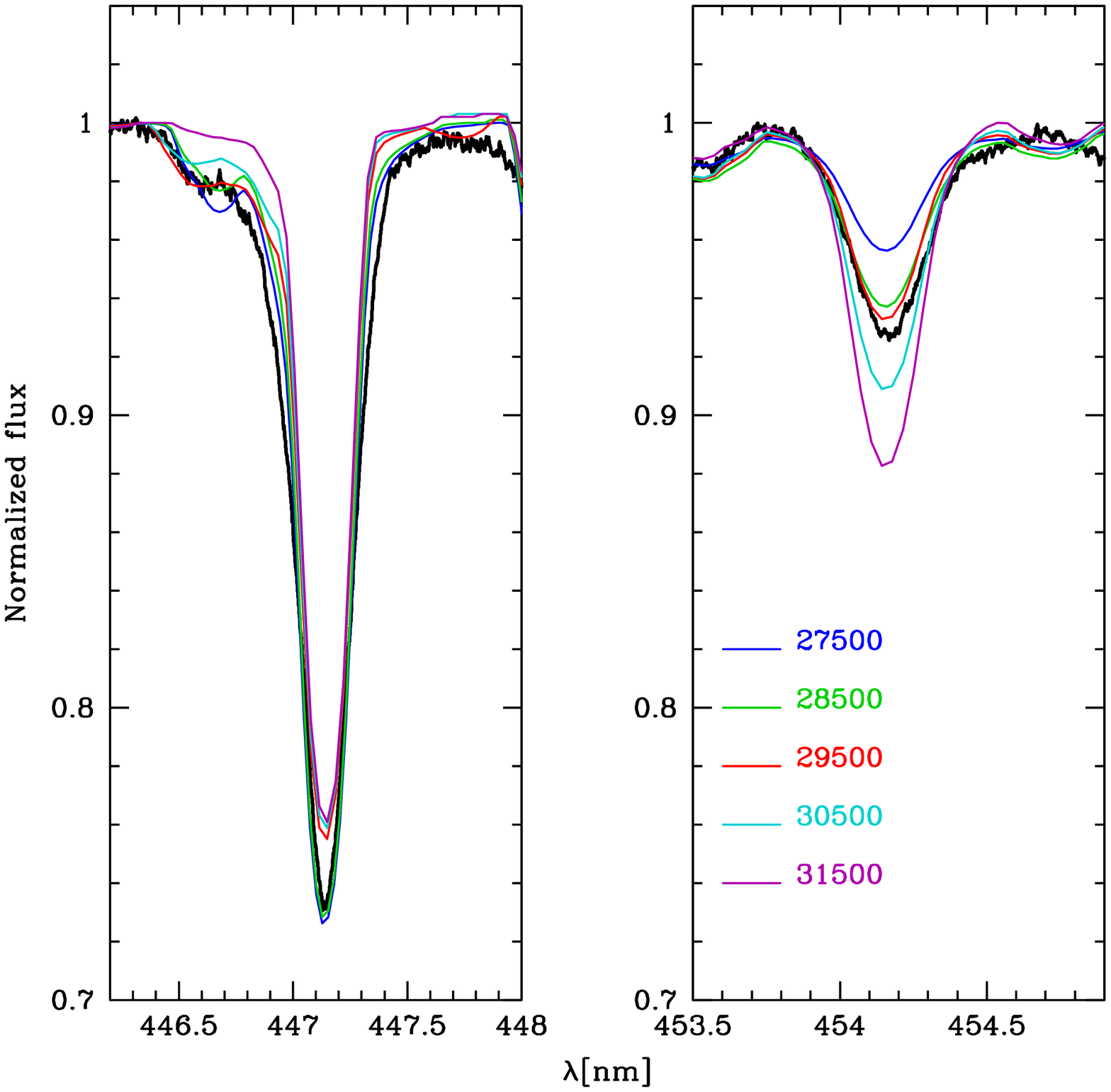}
              \includegraphics[scale=0.45]{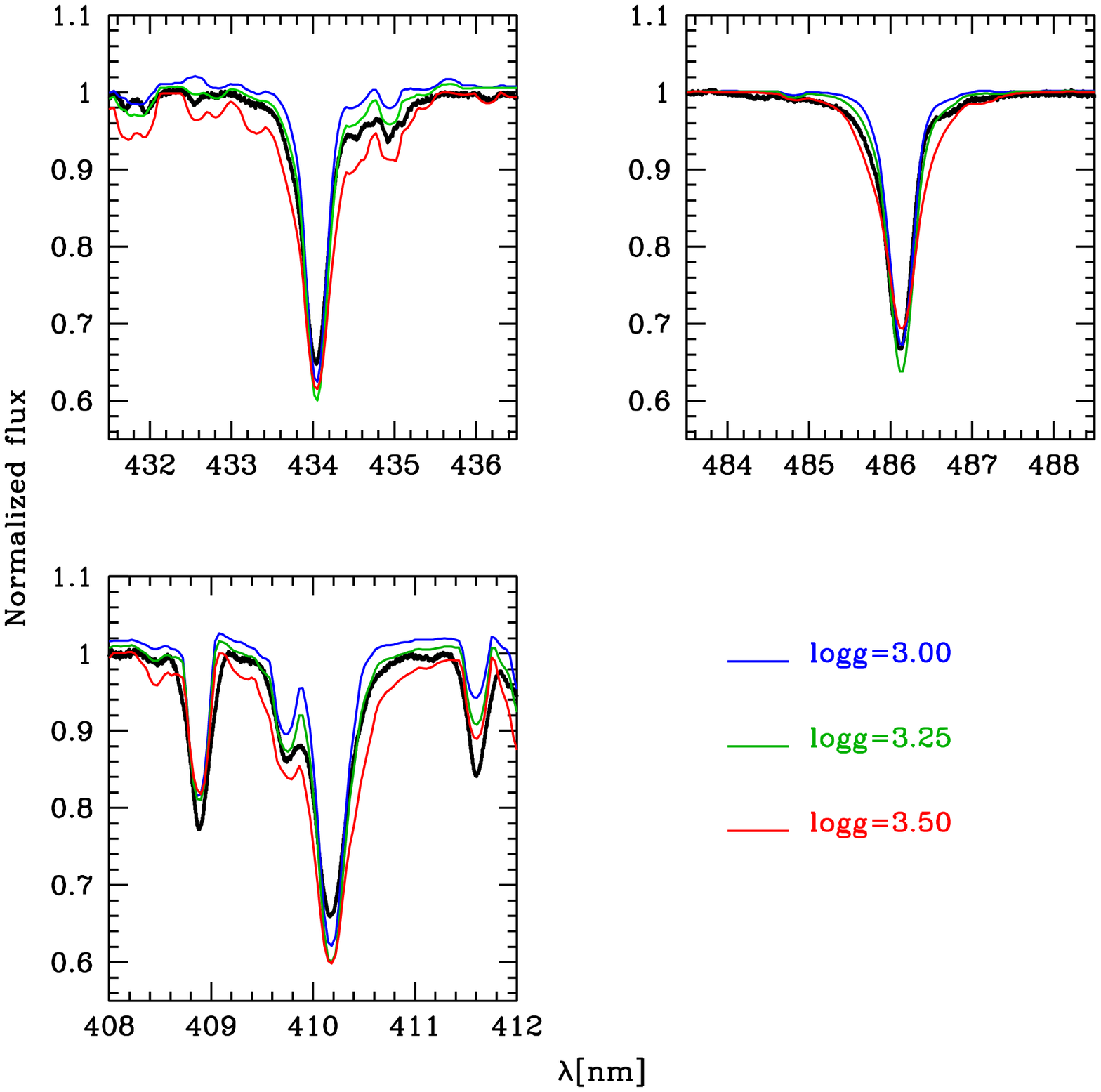}}}  
\caption[]{Modeling the 447~nm He~{\sc i} and the 454~nm He~{\sc ii} lines with atmospheric 
models corresponding to different \teff\ values (2 left panels; \logg\ = 3.25), and the H$\delta$, H$\beta$ 
and H$\epsilon$ lines with models corresponding to different \logg\ values (3 right panels; \teff\ = 29,500 K). For this
plot, we built a mean spectrum out of the spectra obtained on the night of October 18, 2007.} 
\label{fig:tlog}
\end{figure*}

The effective temperature \teff\ was derived from the ratio of He~{\sc ii} to He~{\sc i} 
lines as usually done for O stars.  Comparing with
models from 27,500 to 31,500~K with 1,000~K steps, we find that
$\teff=29,500$~K provides the best fit (see Fig.~\ref{fig:tlog}). 
Given the high quality of the observed spectrum, we can clearly exclude the
27,500 and 31,500~K models.  Models at 28,500 and 30,500~K already
show significant deviations with respect to the best fit.  We thus
(conservatively) conclude that our estimate is accurate to
$\pm$1,000K.  The luminosity was derived from the observed magnitude.
\zetaori\ has $m_{\rm V}=1.76$ and $m_{\rm B}=1.59$ \citep{ma04},
which for an intrinsinc color index (B-V)$_{0}$ of $-0.26$
\citep{mp06}, corresponds to an extinction A$_{\rm V}=0.28$~mag
(assuming R$_{\rm V}$=3.1).  We assumed that the distance to Orion is
equal to $d=414\pm50$~pc \citep[the uncertainty being taken as the
 dispersion among the various recent measurements published in the
 literature,][]{menten07}.  The luminosity of \zetaori\ can thus
be estimated from $m_{\rm V}$, A$_{\rm V}$, $d$ and the bolometric
correction (equal to $-2.73$ for $\teff=29,500$~K, \citealt{mp06}).
We end up with $L=10^{5.64}$~$L_{\odot}$.  Combining the uncertainty
in \teff and $d$ leads to an uncertainty of about 0.15~dex for $L$.
The corresponding radius is thus $R=25\pm5$~\rsun.
We also derived \logg\ from the shape of H$\beta$, H$\delta$ and H$\epsilon$. We 
computed models with \logg\ ranging from 3.0 to 3.75 with 0.25 dex steps. For the three lines, the 
best match is obtained for \logg\ = 3.25. The quantification of the goodness of the fit by means of 
\chisq\ indicates that the uncertainty is of about 0.1~dex.  
From the estimates of \logg\ and $R$, one gets $M=40\pm20$~\msun.

\begin{figure}
\center{\includegraphics[scale=0.43]{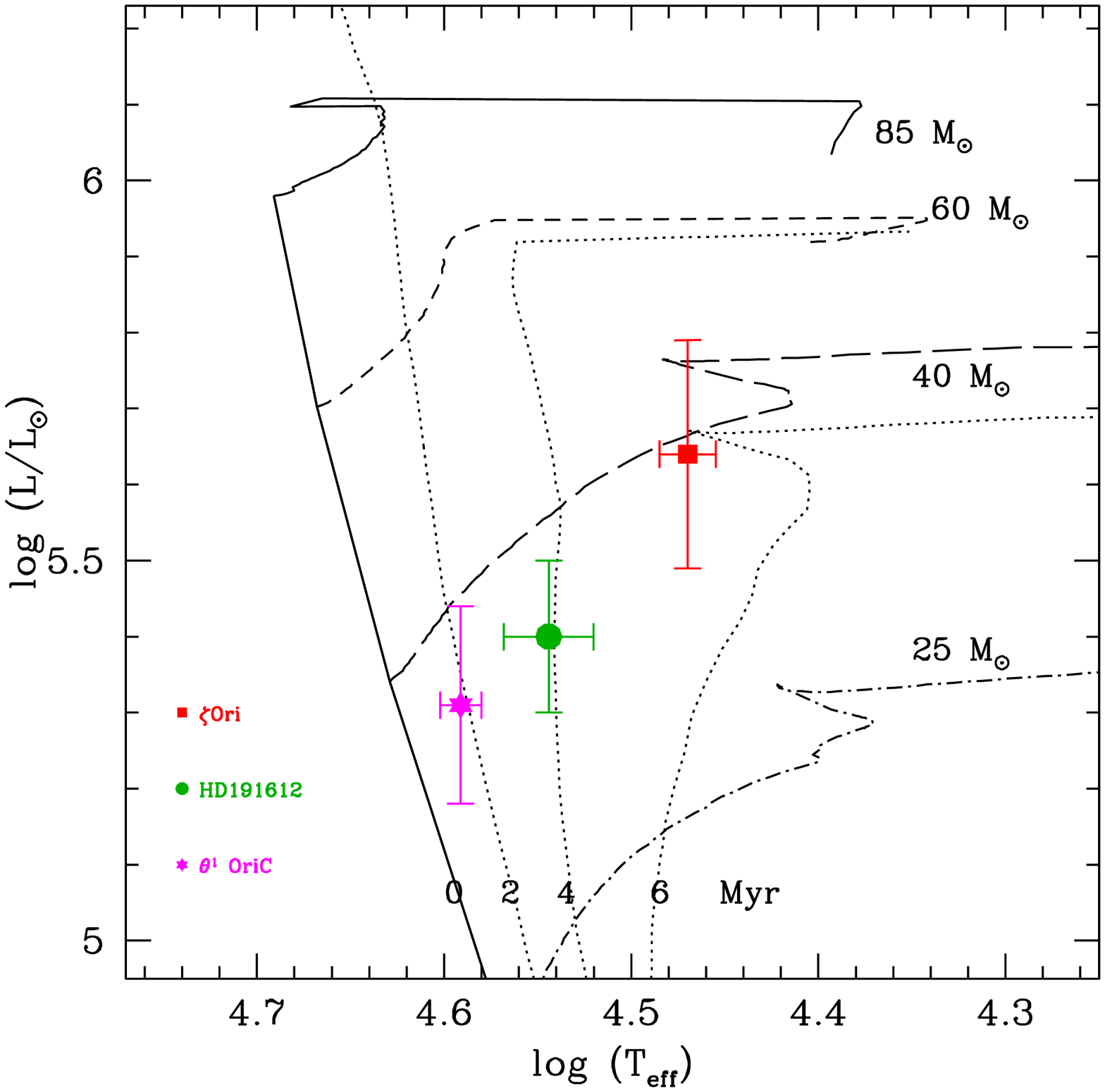}}
\caption[]{HR diagram with the position of \zetaori, \tori\ and HD~191612 indicated. Parameters 
for the last two stars are from \citet{simondiaz06} and \citet{Walborn03} respectively. 
Evolutionary tracks are from \citet{mm03}.  }
\label{fig:hrd}
\end{figure}

In Fig.~\ref{fig:hrd}, we show the position of \zetaori\ in the HR diagram. The two other 
known magnetic O stars (\tori\ and HD~191612) are also reported for comparison. We 
derive an age of  5--6~Myr for \zetaori\ from the Geneva evolutionary tracks \citep{mm03}.  
A simple linear interpolation between the evolutionary 
tracks gives a mass of 39$\pm$8~\msun, in good agreement with (and more accurate than) the 
spectroscopic mass derived above.  At first glance, \zetaori\ thus appears as an evolved 
couterpart of both \tori\ and HD~191612.  

Abundances of a few elements can also be derived. Fig.~\ref{fig:cno} shows such 
determinations for CNO, i.e., the most important elements to constrain stellar 
evolution models. These models predict mixing of CNO-processed material to be more
efficient in fast-rotating stars; accordingly, these stars reveal larger CNO surface abundance
anomalies at an earlier stage. For instance,  \citet{mm03} predict for a solar-composition,
40~\msun\  star with an age of 5~Myr that CNO surface abundances should be equal respectively 
to 0.5, 5 and 0.6 times their original values.  Fig.~\ref{fig:cno} clearly shows that we do not 
observe the large N enrichment. The observed N~{\sc iii} lines are best matched with an
abundance close to the value found in the Orion nebula \citep{esteban04} and in Orion
main-sequence B stars \citep{cunha94}. 
Fitting C~{\sc iii} lines gives uneven results: while some lines are better fitted 
with the Orion nebula abundance, others indicate a small depletion.  Finally, most (but not 
all)  O~{\sc ii} lines indicate that O is underabundant by a factor $\simeq$2. 
The solar composition assumed for all other elements give good fits 
to the observed spectrum (e.g., the 448.0~nm Mg~{\sc ii} and the 
various Si~{\sc iii} and Si~{\sc iv} lines).

\begin{figure*}
\center{\includegraphics[scale=0.55]{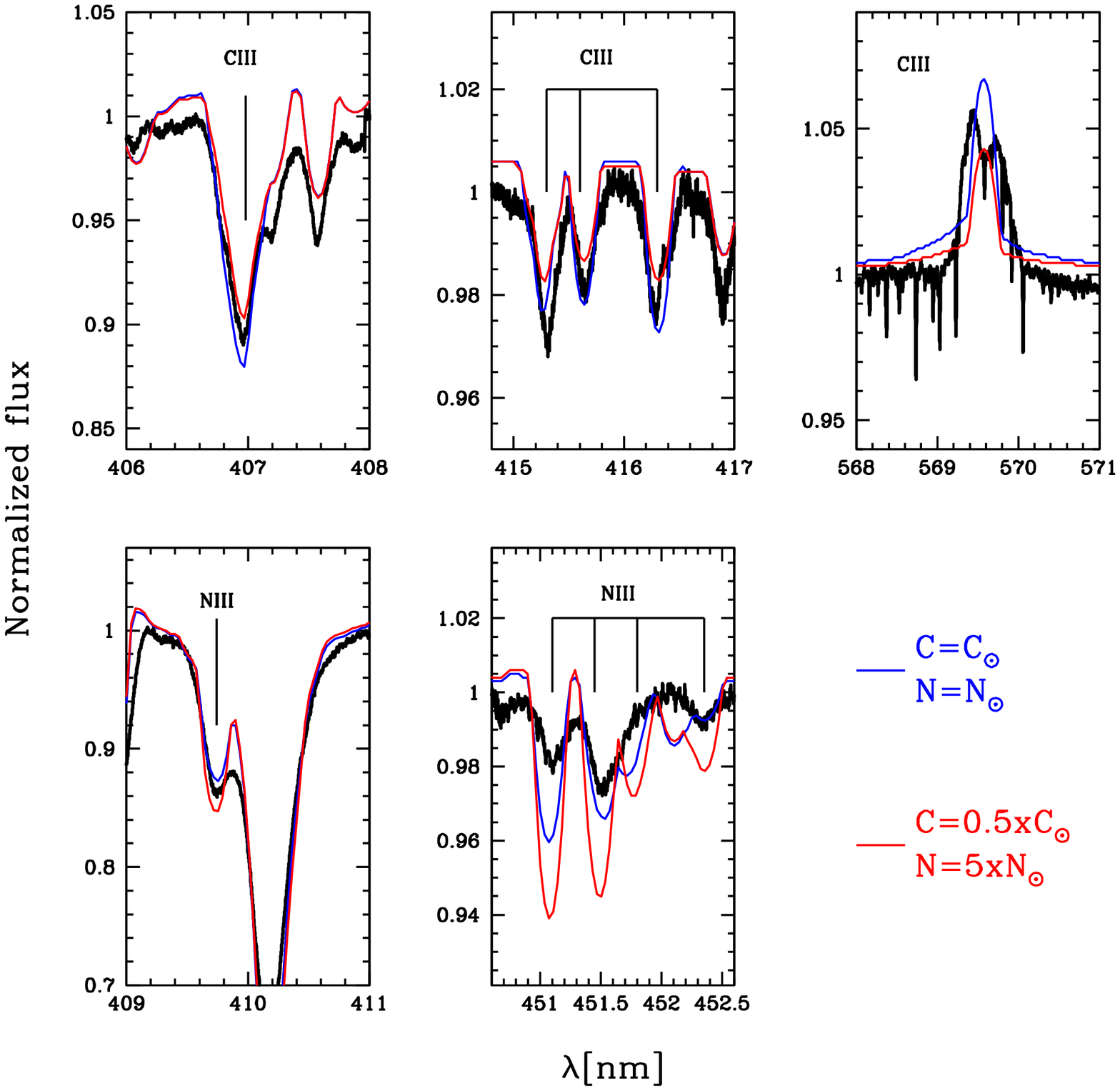}} 
\center{\includegraphics[scale=0.55]{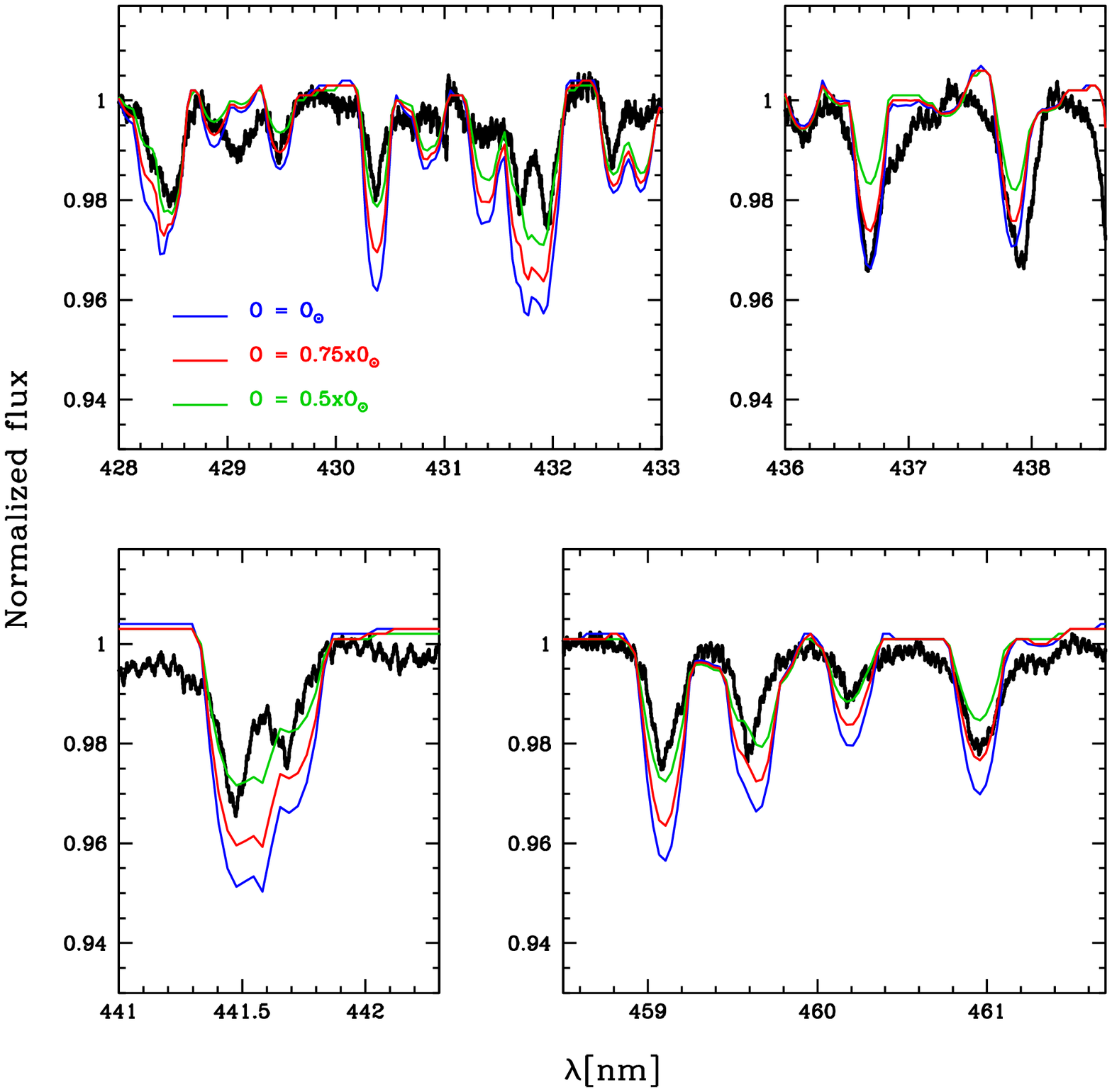}}  
\caption[]{Determining the CN (top) and O (bottom) abundances of \zetaori.  In the top panels, 
the synthetic spectrum corresponding to CN abundances expected for a 40~\msun\ star at an age of 
5~Myr (according to the evolutionary tracks of \citealt{mm03}) is shown in red.  In the bottom
panels, synthetic spectra with O~{\sc ii} lines are shown for models with O = 0.5, 0.75 and 
1.0$~\rm{O}_{\odot}$ (evolutionary models predicting O = 0.65$~\rm{O}_{\odot}$ at 5~Myr).  For this
plot, we built a mean spectrum out of the spectra obtained on the night of October 18, 2007.} 
\label{fig:cno}
\end{figure*}

We used archival IUE spectra to measure the wind terminal velocity from the blueward 
extension of the strong UV P-Cygni profiles and found $\vinf=2,100$~\kms. To estimate 
the mass loss rate, we relied on $H{\alpha}$ only.  
As $H{\alpha}$ is varying with time (see below), we tried to fit the two profiles 
respectively featuring the strongest and weakest emission in the far wings (i.e., the 
part of the profile that can be most reliably fitted with our 1D wind model) to 
derive the range of \Mdot\ over the rotation cycle, yielding 
values of $\Mdot=1.4-1.9\times10^{-6}$~\mspy\ (see Fig.~\ref{fig:mdot}).  
%We stress here that we intend to carry out a full UV-optical analysis using archival IUE data, once the rotation period (see Sect.~\ref{sec:rot}) is better constrained, so that the time of IUE observations can be accurately translated into rotation phases.\\
We achieved a better match to H$\alpha$ adopting a fast velocity law with $\beta=0.8$. 
Fitting the position of H$\alpha$ central absorption requires that clumping starts rather high in the wind (at velocities of about 200 km/s). This is larger than what \citet{bouret05} found for early O supergiants. A summary of our results from this spectroscopic analysis is presented in Table~\ref{tab:spec}.\\ 

\begin{figure}
\center{\includegraphics[scale=0.43]{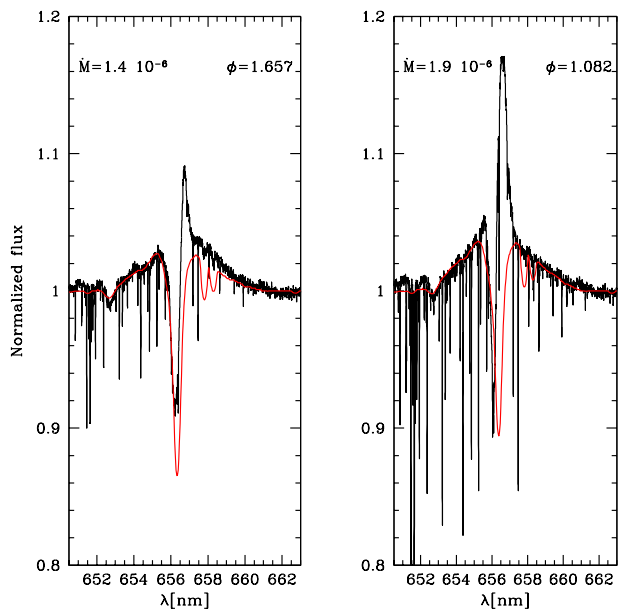}}
\caption[]{Estimating mass loss from H$\alpha$ profiles featuring weakest (left) and strongest 
(right) emission in the far wings, over our 7d campaign.  Matching the broad profile wings 
yields mass loss rates in the range 1.4--1.9$\times10^{-6}$~\mspy.  The two weak lines in the 
red wing are from C~{\sc ii}. } 
\label{fig:mdot}
\end{figure}

Previous determinations of the physical properties of \zetaori\ were made by \citet{ll93}. They gave 
$\teff=30,900$~K, but this estimate was actually based on the effective temperature scale of 
\citet{cg91} and not from a direct analysis of the star with atmospheric 
models\footnote{It is now well established that the \teff\ scale of O stars has been revised 
downward,  \citep[e.g., ][]{msh02, crowther02, repolust04, msh05}};  our finding that $\teff=29,500$~K 
can thus be regarded as a significant improvement with respect to \citet{ll93}.  In the same 
study, the luminosity of \zetaori\ is estimated to $L=10^{5.9}$~$L_{\odot}$, as a result of 
the larger distance they assumed (500~pc) and to a larger \teff\ (and thus a larger bolometric 
correction). Again, our estimate is more robust.

Concerning the abundance patterns, \citet{Raassen08} recently derived solar C and N content as well as a small O depletion from X-ray spectra.  This is consistent with our measurements for these
elements. In particular, their data reveal the absence of strong N enrichment. They also report minor Mg and Si enrichment, but this is not confirmed by our results.  Finally, \citet{ll93} derived a mass loss rate of $2.5\times10^{-6}$~\mspy\ from radio measurements; scaled to a distance of 414~pc (they assumed 500~pc), this corresponds to $\Mdot=1.9\times10^{-6}$~\mspy, in good agreement with our determination. 
We set an upper limit to the actual mass-loss rate of \zetaori\
by searching for the model such that the overall H$\alpha$ profile reaches the peak intensity of
the strongest observed H$\alpha$ emission (corresponding to phase 1.082). We find that 
\Mdot\ $< 2.5\times10^{-6}$~\mspy\ (note that the synthetic profile then strongly overestimates the wings strength). We refer to Sec.~\ref{sec:disc} for a discussion of the origin of the H$\alpha$ emission peak.
Note finally that \citet{ll93} value assumes no wind-clumping ($f=1$), while we adopt $f=0.1$ for the optical study. 
%The strongest H$\alpha$ emission (corresponding to phase 1.082) sets an upper limit to
%the actual mass-loss rate of \zetaori, \Mdot\ $< 2.5\times10^{-6}$~\mspy.
%This is qualitatively consistent with the results of \citet{puls06} concerning the stratification of
%clumping: they found the winds of O stars to be more clumped in the optical formation region than in the outer atmosphere from where the millimeter radition is emitted.

\begin{table}
%\centering 
\caption{Summary of stellar properties of \zetaori, including photospheric and wind parameters derived from the modeling with CMFGEN.  Abundances are expressed relative to Hydrogen. See text for
a discussion of the photospheric abundance patterns relative to the initial/local content.}
%The mass-loss rate is expressed in units of $10^{-6}$ \mspy.}
\begin{tabular}{ll}
\hline
Spectral Type & O9.7 Ib  \\
Distance [pc]   & 414. $\pm 50$ \\
Rotation period [d] & 7.0 $\pm 0.5$ \\
$v \sin i$  [\kms]      & 110. $\pm 10$ \\
$v_{mac}$ [\kms]    & 93. $\pm 9$  \\
Inclination angle $i$ [\degr] & 40. \\
\teff\ [K]             & 29,500 $\pm 1000$ \\
\logg\ [cgs]      & 3.25 $\pm 0.1$ \\ 
log $L$ [$L_\odot$] & 5.64 $\pm 0.15$ \\
M$_{\star}$ [M$_{\odot}$] & 40. $\pm 20$ \\
$\xi_{\rm t}$ [\kms] & 10.  \\
$\dot{M}$ [$\times 10^{-6}$ \mspy] & 1.4 -- 1.9   \\
\vinf\ [\kms]             & 2100.  \\
$\beta$                  & 0.8  \\
$f$                           & 0.1 (default) \\
\vcl\   [\kms]              & 200.  \\
v$_{rad}$  [\kms]   &  45. $\pm 5$ \\
$y=$He/H   &  0.1  \\
C/H       & $2.4 \pm 0.8 \times 10^{-4}$  \\
N/H       & $6.0 \pm 1.8 \times 10^{-5}$  \\
O/H      & $4.6 \pm 1.4 \times 10^{-4}$   \\
  \hline
\end{tabular}
\label{tab:spec}
   \end{table}

\subsection{Temporal variability and rotation}
\label{sec:rot}
Through the Fourier transform of the 580 and 581~nm C~{\sc iv} and O~{\sc iii} 559~nm line profiles 
(averaged over all lines and nights), we can obtain an accurate estimate of the rotational 
broadening parameter \vsini\ (by matching the position of the first zero in the Fourier profile, when this first zero is visible), and of the additional macroscopic turbulent velocity broadening lines (often far) beyond their rotation profiles \citep[e.g.][]{gray81}.  
We find \vsini\ $=110 \pm 10$ \kms (see Fig.~\ref{fig:fft});  for the macroturbulence profile (assumed gaussian), we find $v_{mac} = 93 \pm 9$ \kms (corresponding to a full width at half maximum 
$110 \pm 10$ \kms). 

%Through the Fourier transform of the 580 and 581~nm C~{\sc iv} and O~{\sc iii} line profiles 
%(averaged over all lines and nights), we can obtain an accurate estimate of the rotational 
%broadening parameter \vsini\ by matching the position of the first zero of the Fourier profile 
%(when this zero is visible);  we find \vsini\ to be equal to $110\pm10$~\kms\ (see 
%Fig.~\ref{fig:fft}). 
%{\bf In the Fourier analysis, an additional broadening is needed to describe the line profile.  This is 
%important as it will affect the result from the magnetic analysis; indeed, the amplitude 
%of the modelled Zeeman signature is different for different broadening widths.  
%An additional broadening, the macroturbulence, is needed to describe the line profile. 
%It is assumed here to be gaussian, with a full width at half maximum $v_{mac} = 93$~\kms. It reproduces the observed profile reasonably well (it properly separates $v_{mac}$ from \vsini, thanks to the first lobe in particular), 
%but does not allow to match the slow amplitude decrease of the successive 
%lobes of the Fourier profile (see Fig.~\ref{fig:fft}). 
%A radial-tangential macroturbulence 
%profile may provide a better fit to the data;  this is postponed for a future paper 
%specifically aimed at studying macroturbulence profiles of massive stars.  

\begin{figure}
\center{\includegraphics[scale=0.35,angle=-90]{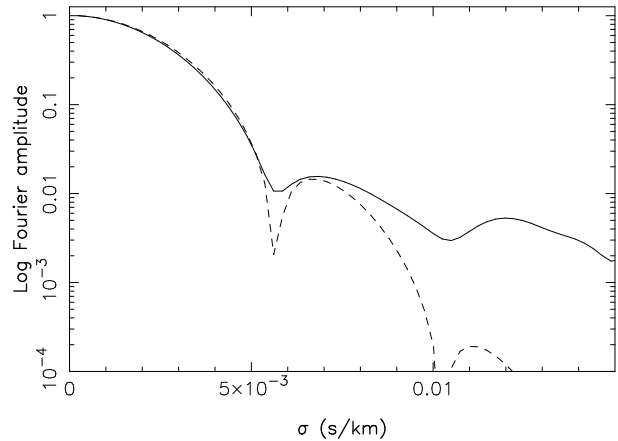}}
\caption[]{Fourier transform of the average 580 and 581~nm C~{\sc iv} and 559~nm 
O~{\sc iii} line profiles of \zetaori. The full line is obtained from the observed profiles, 
the dashed line corresponds to the model.}
\label{fig:fft}
\end{figure}

Given the estimated radius (25~\rsun) and rotational broadening (110~\kms) of \zetaori, 
we conclude that its maximum rotation period $\prot/\sin i$ is equal to 11.5~d.  
Several papers in the refereed literature \citep[e.g.,][]{Kaper97} mention a possible 
rotation period (or half-rotation period) of 6~d for \zetaori\ from variations of 
Balmer lines.  
Looking at how Balmer lines evolve with time during our own observations 
(see Fig.~\ref{fig:bal}, first two left panels) also suggests rotation periods of 6 to 8~d 
depending on which portion of the line profile we focus on.  The central absorption 
of H$\beta$ and H$\alpha$ (both maximum on Oct~18 and Oct~24, i.e.\ cycles 0.66 and 1.51) 
are apparently varying with a timescale of about 6~d, in agreement with the published 
estimate \citep[e.g.,][]{Kaper97}.  
The red emission of H$\alpha$ and the red wing of H$\beta$ (minimum on Oct~18 and Oct~25, 
i.e., cycles 0.66 and 1.66) suggest a slightly longer period.  
The far blue wing of both Balmer lines shows evidence of excess absorption on Oct~19 
and Oct~22 (i.e., cycles 0.82 and 1.23) but not on Oct~25 (cycle 1.66), suggesting a 
(half) period of about 3.5~d;  the far red wing of both lines also show distinct excess 
absorption on Oct~20 (cycle 0.95). 
As a matter of fact, we find that variablity is present in a large majority of lines. 
%It is especially visible in He~{\sc i} 492 nm  or 447 nm lines as well as in the Si~{\sc iv} 409--412 nm  lines. 
Most lines exhibit a clear time variable blueshift, maximum on Oct~19 and Oct~24 (cycles 0.8 and 1.5), i.e. on a timescale of at least 4-5d,  as showed for the case of He~{\sc i} 492 nm Fig.~\ref{fig:bal} (third panel from the left).
%All lines exhibit a blueshift varying with the rotation phase, with maximal amplitude on Oct ~19 and Oct~24 (i.e. at cycles 0.8 and 1.5 respectively), as showed for the case of He~{\sc i} 492 nm Fig.~\ref{fig:bal} (third panel from the left). 
The average blueshift (with respect to C~{\sc iv} 580--581 nm) is $-16$ \kms, while the peak-to-peak maximal amplitude is $-17$ \kms,  well below the radial velocity variations induced by the companion of \zetaori\ \citep{hummel00}.

\begin{figure*}
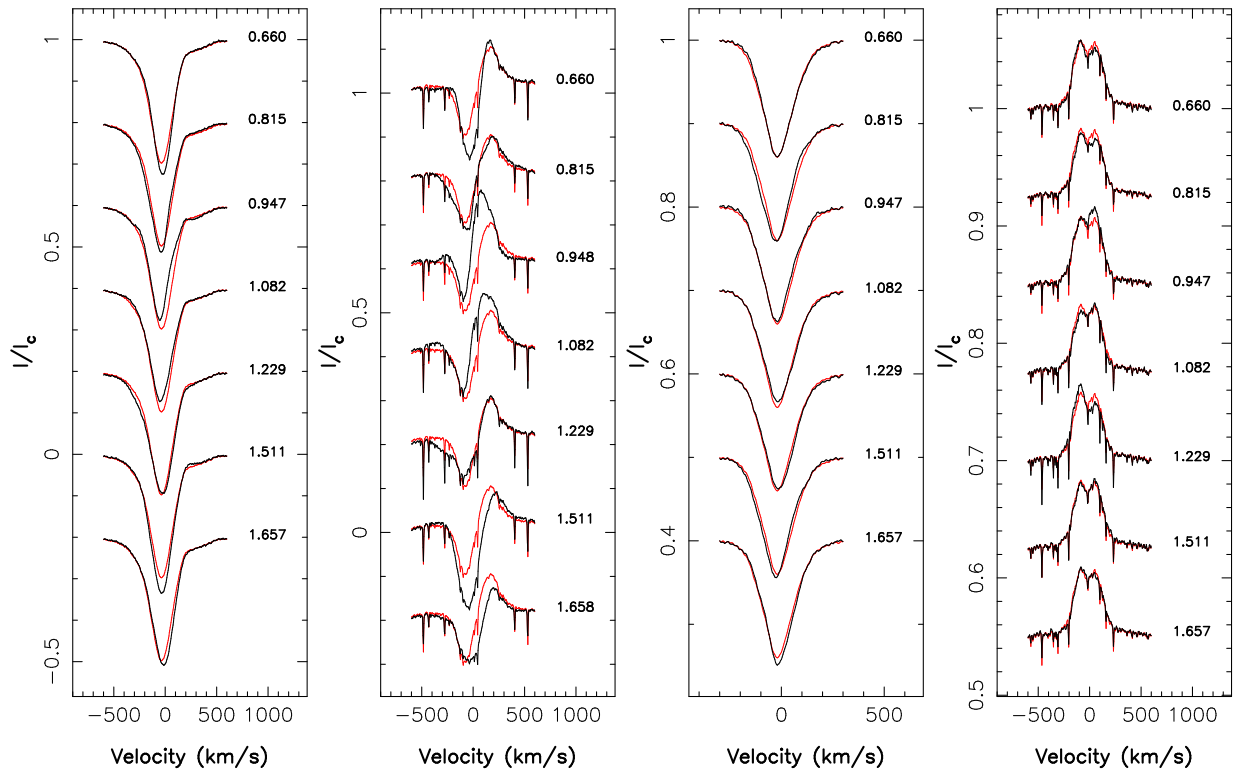

\center{\includegraphics[scale=0.55,angle=-90]{fig/bouret_zetaori_fig7a.eps}
              \includegraphics[scale=0.55,angle=-90]{fig/bouret_zetaori_fig7b.eps}
              \includegraphics[scale=0.55,angle=-90]{fig/bouret_zetaori_fig7c.eps}
              \includegraphics[scale=0.55,angle=-90]{fig/bouret_zetaori_fig7d.eps}
              }
\caption[]{Temporal variations of the H$\beta$ (left), H$\alpha$  (second from left), 
He~{\sc i} 492~nm (third from left) and 569.59~nm C~{\sc iii} (right) lines throughout our observing run.  
The time-averaged profile is plotted in red to emphasise variations.  
The rotational cycle of each observation (assuming a rotation period of 7~d) 
is written next to each profile.  All profiles are shown in the star's rest frame
(i.e., shifted by 45~\kms\ with respect to the heliocentric rest frame).  } 
\label{fig:bal}
\end{figure*}

We also detect temporal variations in the 569.59~nm C~{\sc iii} double-peak emission line 
(see Fig.~\ref{fig:bal}, right panel) where the relative intensity of both peaks vary from one night 
to the next.  In our observations, the red peak features maximum emission on Oct~20 and Oct~23 
(cycles 0.95 and 1.51) while the blue peak shows maximum emission on Oct~22 (cycle 1.23), apparently 
in antiphase with the red peak of the same line and in phase with the excess absorption episode 
occuring in the far blue wing of both Balmer lines.  The intensity ratio of both peaks is 
varying on a timescale of about 3--4~d and may potentially be a good indicator of the 
rotation half-period.  

Given that we detect clear modulation on a period of about 3.5~d both in the far blue wings of 
both Balmer lines and in emission peak intensity ratio of the C~{\sc iii} line, we interpret 
the $\simeq$7~d period as the rotation period (rather than half the period as 
\citealt{Kaper97}). The 4-5d timescale seen in most photospheric lines is potentially also 
compatible with a 7d rotation period (e.g. if the two maximum blueshifts are unevenly spaced in rotation phase). 

We therefore assume in the following a 7~d timescale to phase our data, 
and use the following ephemeris to compute rotational phases:  
\begin{equation}
\mbox{HJD} = 2454380.0 + 7.0 E.
\label{eq:eph}
\end{equation}
We further discuss below the determination of the rotation period, using the detected Zeeman 
signatures to probe rotational modulation.  
Assuming a rotation period of about 7~d implies that the star is seen at an 
inclination (with respect to the rotation axis) of about 40\degr, and that the equatorial 
velocity of \zetaori\ is about 170~\kms.  While this is not quite extreme rotation by 
O star standards, this is already much higher than the two magnetic stars known to date 
(\tori\ and HD~191612, both featuring equatorial rotation velocities lower than 30~\kms).

\section{Modelling the magnetic topology of \zetaori}
\label{sec:magt}

To model the Zeeman signatures of \zetaori, we use the imaging code of 
\citet{Donati06b}.  The magnetic topology at the surface of the star is 
reconstructed as a spherical-harmonic expansion, whose coefficients are 
adjusted (with a maximum-entropy image reconstruction code) to ensure 
that the synthetic Zeeman signatures corresponding to the reconstructed 
magnetic topology match the observed ones at noise level.  The magnetic 
image that we derive can thus be regarded as the simplest topology compatible 
with the data.  

This new imaging method has several advantages with respect to the older 
one of \citet{Brown91} and \citet{Donati97b}.  The reconstructed field 
is directly expressed as the sum of a poloidal and a toroidal 
field.  Moreover, we have a direct and obvious way of constraining the 
degree of complexity of the reconstructed field topology by limiting the 
spherical harmonic expansion at a given maximum $\ell$ value, depending on 
the quality and temporal sampling of the available Zeeman data.  

For this modeling, we assume that $\vsini=110$~\kms\ and $i=40$\degr, 
as derived in Sec.~\ref{sec:rot}.  We also assume that the star and its 
magnetic topology are rotating as a solid body with a rotation period of 7~d;  
different values of the rotation period are also used to evaluate how much 
the result we obtain is sensitive to this parameter (see below).  The line 
profile model (including macroturbulence broadening) used to describe the 
synthetic Zeeman signatures is the same as that introduced in Sec.~\ref{sec:spec} 
to describe the observed photospheric lines, i.e., a simple gaussian at an 
average wavelength of 500~nm and with full width at half maximum equal to 
110~\kms.  Zeeman signatures are obtained by assuming the weak field 
approximation and an average Land\'e factor of 1.1.  

The complete set of Zeeman signatures and their corresponding null profiles 
are shown in Fig.~\ref{fig:fit} along with the maximum entropy fit to the 
data, assuming either a simple dipole field or a more complex magnetic 
geometry (limited to $\ell=3$).  The reduced \chinu\ associated 
to the global set of $V$ and $N$ profiles with respect to a non magnetic ($B=0$) 
model are equal to 1.25 and 0.99 respectively (for a total number of 392 data points), 
indicating that the magnetic signal in the $V$ profiles is unambiguously detected 
at a 10$\sigma$ level while no significant signal is observed in the $N$ profiles.  

\begin{figure}
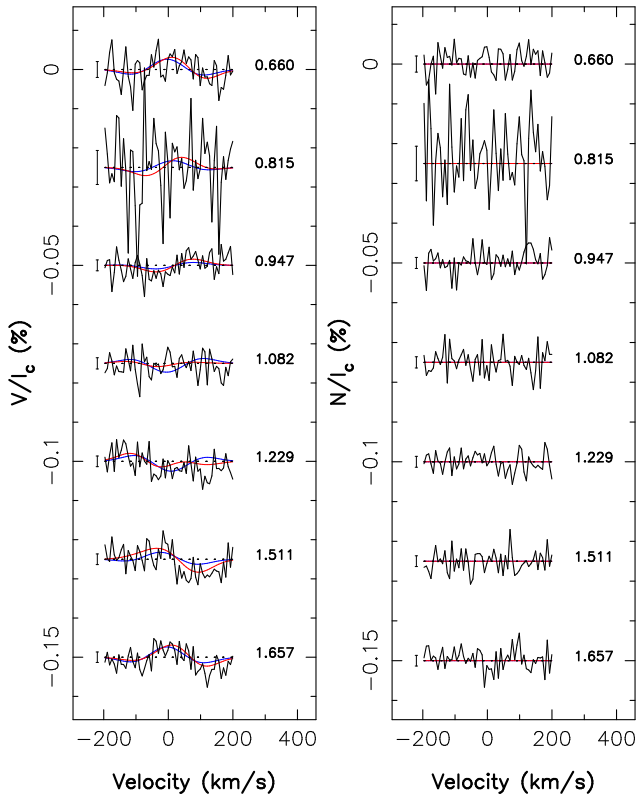

\center{\includegraphics[scale=0.57,angle=-90]{fig/bouret_zetaori_fig8a.eps}
        \includegraphics[scale=0.57,angle=-90]{fig/bouret_zetaori_fig8b.eps}}
\caption[]{Observed (black) and modeled (red \& blue) Stokes $V$ signatures 
(left) and null $N$ profiles (right) of \zetaori.  The blue line corresponds to a 
simple dipole, while the red one corresponds to a more complex field having 
$\ell=3$.  A clear Zeeman signal is detected and consistently modeled in the 
Stokes $V$ profiles. The rotational cycle of each observation (assuming a 
rotation period of 7~d) is written next to each profile.  A 1$\sigma$ error
bar is also plotted left to each profile. } 
\label{fig:fit}
\end{figure}

Assuming that the star hosts a simple dipole field provides a better fit to the 
data;  however, the resulting \chinu\ (equal to 1.14) is still significantly 
larger than 1, indicating that the magnetic topology of \zetaori\ is likely 
more complex.  Using a spherical harmonics series expanded up to $\ell=3$ 
provides a unit \chinu\ fit to the data, i.e., is successfull at reproducing
the data down to noise level;  in particular, it provides a much better fit to 
the data at cycle 1.51 on the red side (negative dip) of the line profile 
(see Fig.~\ref{fig:fit}).  Unsurprisingly, carrying out the same analysis on 
the $N$ spectra only yields flat synthetic profiles, the corresponding \chinu\ 
for a non-magnetic model being already below 1.  

\begin{figure*}
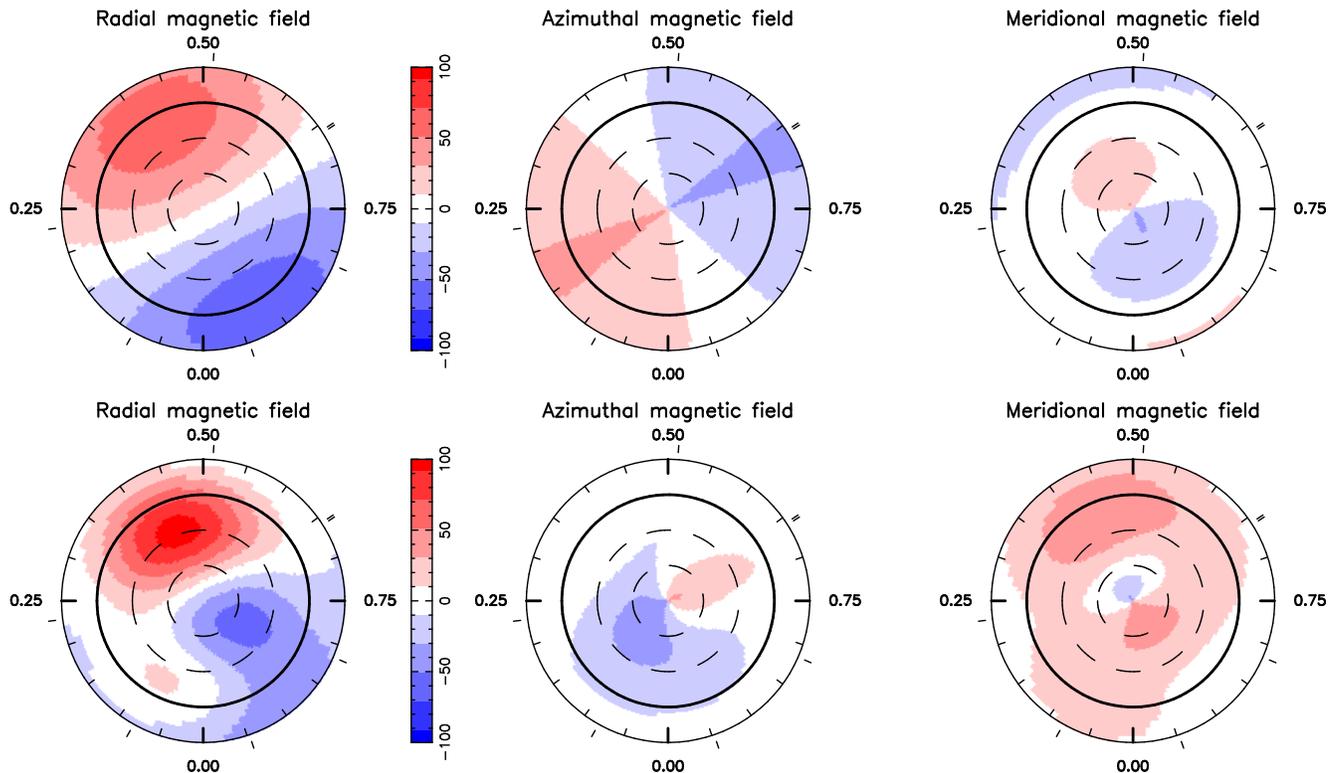

\center{\includegraphics[scale=0.73]{fig/bouret_zetaori_fig9a.eps}} 
\center{\includegraphics[scale=0.73]{fig/bouret_zetaori_fig9b.eps}}
\caption[]{Reconstructed magnetic topology of \zetaori, assuming either a dipole 
magnetic field (top) or a more complex field having $\ell=3$ (bottom).  Only the 
second (lower) topology provides a $\chinu=1$ fit to the Stokes $V$ data.  In both 
cases, the three field components are displayed from left to right (flux values 
labelled in G).  The star is shown in flattened polar projection down to latitudes
of $-30\degr$, with the equator depicted as a bold circle and parallels as dashed 
circles.  Radial ticks around each plot indicate phases of observations.  } 
\label{fig:map}
\end{figure*}

The dipolar and $\ell=3$ magnetic topologies derived from the data are both 
shown in Fig.~\ref{fig:map}.  The reconstructed dipole has a strength of $61\pm10$~G, 
is roughly perpendicular to the rotation axis (inclination angle $\beta=83\pm10$\degr) 
with the positive pole facing the observer at phase $0.42\pm0.03$.  The second (more 
complex) magnetic topology shows more concentrated features (where the field reaches 
as much 100~G) and is mainly poloidal (the toroidal component containing less than 5\% 
of the reconstructed magnetic energy).  Given the limited 
resolution we have access to on the star (about six resolution elements across the 
equator given the fairly large width of the local profile), the moderate accuracy 
to which the Zeeman signatures are detected and the moderate phase coverage of 
the collected data, there is no real point at carrying out reconstructions with even 
more complex field topologies.  

We carried out the same analysis for different values of the rotation period 
(and corresponding inclination), looking for the period that 
produces the magnetic image with the smallest information content and a $\chinu=1$ fit 
to the data.  From this process, we derive that 7.5~d is a marginally better rotation 
period, with a 1$\sigma$ error bar equal to about 0.5~d.  The magnetic topologies 
derived for this value of the rotation period are very similar to those of 
Fig.~\ref{fig:map}.  Using this criterion, we also find that periods of 12 to 14~d 
are less likely to be the true rotation period of \zetaori.  Note that this constraint 
on the rotation period depends on the assumed magnetic geometry (here limited 
to an $\ell\leq3$ spherical harmonic expansion) given the limited span of our 
observations (7~d).

\section{Discussion}
\label{sec:disc}

We report in this paper the detection and the first modeling attempt of the weak large-scale 
magnetic field of the O9.7 supergiant \zetaori.  We detect a field that corresponds to local 
surface magnetic fluxes of only a few tens of G. The field is everywhere lower than 100~G, 
making it (by far) the weakest magnetic field ever reported in a hot massive star 
\citep{Donati90, auriere07}.  In particular, this magnetic field is weaker than the 
thermal equipartition limit, equal to about 100~G for \zetaori;  this is the first 
sub-equipartition field unambiguously detected in a hot star. The magnetic chemically 
peculiar stars all show fields larger than their thermal equipartition limit 
\citep{auriere07}.  This detection also brings the number of known magnetic O stars to three, with 
\zetaori\ thus joining \tori\ \citep{Donati02} and HD~191612 \citep{Donati06a}.  
This is also the first magnetic detection in a 'normal' rapidly rotating O star.  

The detailed spectral modeling of \zetaori\ provides
$\teff=29,500\pm1,000$~K and $\logg=3.25\pm0.10$ with normal
abundances.  It follows that \zetaori\ is a 40~\msun\ star with
a radius equal to about 25~\rsun, seen from the Earth at an
inclination angle of 40\degr.  With an age of about 5--6~Myr,
\zetaori\ essentially appears as an evolved counterpart of both \tori\ and
HD~191612.  Given its evolutionary stage, \zetaori\ is expected to
show significant N enrichment at its surface (as well as moderate C
and O depletion); the normal nitrogen abundance that we measure is thus
surprising.  It is tempting to suggest that magnetic fields may play a
role in this process; this is however not what the first evolutionary
models including magnetic field predict \citep{Maeder05}.  More work
(both on the observational and theoretical side) are required to
investigate this issue further.  From a fit to H$\alpha$, we estimate 
that the mass-loss rate is about 1.4--1.9$\times10^{-6}$~\mspy.  
From the temporal variability of spectral lines and the modulation of Zeeman signatures, we 
find that the period of \zetaori\ is about 7d. This is compatible with the 
\vsini\ that we measure (from the Fourier shape of the photospheric C~{\sc iv} lines) 
and the radius that we derive (from the spectral synthesis), provided the star 
is view at intermediate inclinations ($i=40$ \degr).

Given that \zetaori\ is typically 3 and 1.4 times larger in 
size than \tori\ and HD~191612 respectively, we find that its overall unsigned magnetic 
flux (i.e., the integral of the absolute value of the magnetic field over stellar surface)
is slightly larger (by a factor of about 1.5) than that of \tori\ but much smaller 
than that of HD~191612 (by about an order of magnitude).  

The extremely long rotation period of HD~191612 (about 538~d) suggests that the magnetic 
field is likely responsible for having dissipated (through confined mass loss) most of 
the angular momentum of HD~191612 \citep{Donati06a}.  The slow rotation rate and 
extreme youth of \tori\ also suggests that primordial magnetic fields pervading the 
parent molecular cloud must have a strong impact onto the angular momentum dissipation 
throughout the cloud collapse, in qualitative agreement with what numerical simulations 
predict \citep[e.g.,][]{Hennebelle08a}.  In this context, one 
would expect \zetaori\ to rotate, if not as slowly as HD~191612 (whose intrinsic 
magnetic flux is much higher), at least more slowly than \tori\ (whose intrinsic 
magnetic flux is similar) given its later evolution stage;  this is however not what 
we observe.  No more than speculations can be proposed at this stage.  
One possibility is that the magnetic field of \zetaori\ is not of fossil origin (as 
opposed to that of \tori\ and HD~191612) but rather dynamo generated, making 
the rotational evolution of \zetaori\ and \tori\ hardly comparable.  
The detected magnetic field is indeed much weaker than the critical limit above 
which MHD instabilities are inhibited (about six times the equipartition field or 
600~G in the case of \zetaori, \citealt{auriere07}) and may thus result from exotic 
dynamo action;  the non-dipolar nature of the detected field could be additional 
evidence in favour of this interpretation, fossil fields being expected to have 
very simple topologies in evolved stars.  
Additional spectropolarimetric observations of \zetaori\ at different epochs 
(searching for potential variability of the large-scale field) and of similar 
'normal' rapidly-rotating stars are obviously necessary to explore this issue 
in more details.  

Computing the wind magnetic confinement parameter $\eta_{\ast}$ of
\citet{udDoula02} for \zetaori\ and taking $B\simeq30-50$~G (at the
magnetic equator), $R=25$~\rsun, $\Mdot=2\times10^{-6}$~\mspy\ and
$\vinf=2,100$~\kms\ (see Secs.~\ref{sec:spec} and \ref{sec:magt})
yields $\eta_{\ast}\simeq0.03-0.07$.  The magnetic field of
\zetaori\ is therefore just strong enough (according to theoretical
predictions) to start distorting the wind significantly
\citep{udDoula02}.  The observed rotational modulation in H$\alpha$,
H$\beta$ and the C~{\sc iii} lines confirms this first conclusion;  
the variation in mass loss rate that we measure, corresponding to a 
density contrast of $\simeq$1.4, is compatible with what numerical simulations 
of magnetically confined winds predict \citep[see, e.g., Fig~8 of][]{udDoula02}.  

Note also that the observed line blueshift and asymmetries (cf. Sec.~\ref{sec:rot} and 
Fig.~\ref{fig:bal} for an illustration on the case of He~{\sc i} 492~nm) are (rotation) phase dependant.
The observed blueshift of most lines is maximum when the magnetic poles (i.e. the open 
field lines) cross the line of sight (at phase 0.8 and 0.45, see Fig.~\ref{fig:bal});   
more data (collected in particular over a longer baseline and densely sampling the rotation cycle) are of course needed to confirm this and to specify how exactly the line shifts and shape are varying with rotation phase (e.g. with 2 unevenly spaced maxima in the line blueshift per rotation period). 
It however suggests already ${\it (i)}$ that the line profile variations reflect the varying conditions in which the wind form at the surface of \zetaori\ (as a result of the varying local field topology over the star) and ${\it (ii)}$ that these variations can potentially be used to trace the density at the base of the wind and its variations with the local magnetic topology over the surface of the star. 
%and reach a maximum when magnetic poles are on the line-of-sight. This suggests that
%these variation of blueshift/asymmetries are related to variations of the wind confinment,
%the latter varying with the magnetic configuration. A careful investigation (based on data
%sampling long enough periods) would in principle allow to model precisely the densitiy
%stratification and velocity law at the base of the wind, in regions where the field intensity 
%is strong(and vertical, near the poles) and in other regions where its intensity is weaker 
%(or where the field is horizontal).  

On both \tori\ and HD~191612, H$\alpha$ is rotationally modulated as a result of the 
magnetic obliquity (with respect to the rotation axis), with maximum emission occuring 
when the magnetic pole comes closest to the observer.  Similarly, maximum absorption 
in UV lines (with highest blueshifted velocities) are observed when the magnetic equator 
is crossing the line of sight.  
Extrapolating these results to \zetaori, we would have first expected H$\alpha$ in \zetaori\ 
to show maximum emission twice per rotation cycle, at phases of about 0.40 and 0.90 (see 
Fig.~\ref{fig:map}), in contradiction with what we observe;  while Balmer emission indeed 
peaks at phase 0.95, phase 0.51 rather corresponds to minimum (rather than maximum) emission 
(see Fig.~\ref{fig:bal}).  

The analogy with \tori\ and HD~191612 can obviously not be directly applied to \zetaori.  
Given the much weaker wind magnetic confinement parameter of \zetaori\ (roughly equal to 
10 for both \tori\ and HD~191612), this is not altogether very surprising.  In particular, 
the Alfven radius is much closer to the surface of the star in \zetaori, probably not 
further than 0.05--0.1~\rstar\ above the surface\footnote{The corotation radius, i.e., the 
radius at which the Keplerian period equals the rotation period at the surface of the star, 
is equal to about 2~\rstar\ in \zetaori, i.e., 1~\rstar\ above the surface of the star.} 
(as opposed to 1~\rstar\ above the surface 
for \tori, \citealt{Donati02}).  In the magnetically confined wind-shock model 
\citep{bm97, Donati02}, the rotational modulation of H$\alpha$ emission can be ascribed, 
in a generic way, to the varying aspect of the magnetic equatorial disc up to the Alfv\'en 
radius;  in the case of \zetaori, this variation is expected to be minimal.  We speculate 
that most of the redshifted H$\alpha$ emission comes from a region located just above the 
photosphere (at the very base of the wind) and essentially reflects a difference between 
both magnetic poles (strongest emission being observed in conjunction with the weakest 
magnetic pole, see Fig.~~\ref{fig:map}, bottom panel).  

We also note that the excess absorption that both Balmer lines exhibit twice per 
rotation period in their distant blue wing (see Fig.~\ref{fig:bal}, left and right panels) 
behave as UV absorption lines do in \tori;  we propose that they reflect the magnetic 
equator crossing the line of sight (at rotation phases of about 0.20 and 0.75).  The 
maximum radial velocities associated to these absorption components (up to about 500~\kms, 
i.e., less than 0.25\vinf) confirm that they correspond to material located within the 
Alfv\'en radius.  More data (densely sampled over several rotation cycles) are needed 
to investigate this issue more closely, and to pin down unambiguously the origin of the 
various H$\alpha$ and H$\beta$ components.  

The 569.6~nm C~{\sc iii} double-peak emission line is also a significant difference with 
respect to \tori\ and HD~191612 (where the line only features a single peak emission).  
The observed modulation is apparently related to the magnetic topology, with 
the red emission peaking at phases of maximum magnetic field and the blue emission peaking 
when the magnetic equator is crossing the line of sight (both phenomena occuring twice per 
rotation period).  The maximum velocities of both components (up to about 200~\kms) also 
argue for the formation of this line within the Alfv\'en radius and, therefore, it is likely a 
good indicator of the influence of the magnetic field on the launching of the wind. Further 
observational and theoretical studies are again required to examine how this line responds 
to a magnetised wind.  

At the very least, our results demonstrate that the magnetic field of \zetaori\ has a 
significant impact on the wind despite being below pressure-equipartition and the 
weakest detected ever in a hot star.  Given the obvious 
importance of this result for our understanding of massive magnetic stars, we 
need to confirm and expand the present analysis with new data collected over several 
rotation periods of \zetaori, i.e., over a minimum of 20 nights;  renewed observations 
will indeed allow us (i) to obtain an accurate measurement of the rotation period, (ii) to derive 
a fully reliable modeling of the large-scale magnetic topology and (iii) to estimate whether 
the field is  intrinsically variable as usual for dynamo topologies, e.g., on a typical timescale 
of 1~yr, and (iv) a detailed 
account of how wind lines (and in particular H$\alpha$, H$\beta$ and the 569.6~nm 
C~{\sc iii} lines) are modulated with the viewing aspect of the magnetic topology.  

\section*{Acknowledgments} 
{We thank our referee, O. Stahl, for valuable comments. Thanks to John Hillier for constant support with his code CMFGEN. FM acknowldeges generous allocation of computing time from the CINES. JCB acknowledges financial support from the French National Research Agency (ANR) through program number ANR-06-BLAN-0105.}

%% Bibliography

\bibliography{bouret_zetaori}

\bibliographystyle{mn2e}

\end{document}